\documentstyle[mprocl,epsf]{article}

\def\Journal#1#2#3#4{{#1} {\bf #2}, #3 (#4)}

\def\PRD{{\em Phys. Rev.} D}

\begin{document}

\title{SINGULARITIES INSIDE HAIRY BLACK HOLES}

\author{D.V. GAL'TSOV}

\address{Department of Theoretical Physics, \\
        Moscow State University, 119899 Moscow, Russia}

\author{E.E. DONETS}  

\address{Laboratory of High Energies, JINR, 141980 Dubna, Russia}
                             
\author{M.Yu. ZOTOV}  

\address{Skobeltsyn Institute of Nuclear Physics, \\
        Moscow State University, 119899 Moscow, Russia}

\maketitle\abstracts{We show that the Strong Cosmic Censorship
is supported by the behavior of generic solutions 
on the class of static spherically symmetric black holes 
in gravitating gauge models and their stringy generalizations.}

The validity of the Strong Cosmic Censorship hypothesis
(SCC) which forbids locally naked, i.e., timelike, singularities
is not quite clear. Destruction of the internal Cauchy horizon in charged
black holes due to the mass inflation phenomenon is likely to support
the SCC conjecture, but these considerations are mostly perturbative,
and the final answer about the nature of singularity in mass inflation
is not settled yet. Both analytical and numerical analysis is rather
complicated as far as a dynamical collapse picture is concerned. Meanwhile
certain conclusions may be deduced already from a simpler analysis
of the static solutions. Here we suggest to use an argument of
the generic solutions behavior
to judge about the validity of the SCC in the framework of gravity coupled
gauge theories and their stringy generalizations. Black holes in such
a framework may have different types of singularities both satisfying
and violating SCC. However the latter can be shown to form only 
a set of zero measure in the parameter space.

By analogy with Schwarzschild (S) and Reissner--Nordstr\"om (RN) solutions,
one could expect singularities of either spacelike or  timelike
nature with the divergence of the curvature invariants
being given by some inverse powers of the radius of two-spheres 
(power-law singularities). It turns out that generic singularities 
in the non-Abelian static black holes 
in the presence of scalar fields (such as Higgs or dilaton) are also
of power-law type, although, contrary to  S and RN cases,
with some non-integer power indices depending
on the black hole mass. In the absence of scalar fields, i.e., in the
pure Einstein--Yang--Mills (EYM) theory, the generic singularity 
is drastically different and has an oscillating nature like the BKL 
singularity in the Bianchi IX cosmology. The $\Psi _2$ component of the Weyl 
tensor is infinitely oscillating with growing amplitude, while the metric
does not develop internal Cauchy horizons. The interior region of such
a black hole is isometric to the closed Kantowski--Sachs (KS)
cosmology which does not belong to the Bianchi types, thus the nature
of oscillations is rather different from that in the BKL case. 

Consider first the pure EYM $SU(2)$ theory (see~\cite{dgz} for notation and
conventions).  The metric is chosen in the form 
\[
 ds^2 = (\Delta/r^2) \sigma^2 dt^2 - (r^2 / \Delta) dr^2 - r^2 d \Omega^2 ,
\]
while the YM field is fully described by the function $W(r)$.
The field equations consist of a coupled system for $W$ and the 
mass function $m(r)$, $\Delta=r^2 - 2 m r$:
\begin{equation}\label{eq1}
  \Delta (W'/r)' + F W' = W V/r , \quad  m'= 4 \pi r^2 \epsilon , \label{eq2}
\end{equation}
 where $V=(W^2-1)$, $F=1 - V^2/r^2$, and a decoupled equation for $\sigma$:
\[
 (\ln \sigma)' = 4 \pi r^3 |\Delta|^{-1} (\epsilon + p_r),
\]
where the energy density and the radial pressure are
 \begin{equation} \label{ep}
 4 \pi \epsilon = r^{-4}\left( |\Delta| W'^2 + V^2/2 \right), \quad
 4 \pi p_r = r^{-4}\left( |\Delta| W'^2 - V^2/2 \right).  
\end{equation}
These formulas are valid both outside ($\Delta>0$) and inside ($\Delta<0$)
the black hole. Note that the radial pressure is not positively definite.
Tangential pressure $p_t = (\epsilon - p_r)/2$ is strictly positive
due to the tracelessness of the stress-tensor of the YM field.

If one starts looking for the local solutions around $r=0$ in terms of 
power-law expansions, then one finds one solution describing a timelike 
singularity:
\begin{equation} \label{rn}
 W = W_0 - W_0 r^2/(2 V_0) + c r^3 + O(r^4), \quad
 \Delta = V_0^2  - 2 m_0 r + r^2 + O(r^3),
\end{equation}
where $W(0)= W_0\ne \pm 1$. This three--parameter ($W_0$, $m_0$, $c$)
solution corresponds to the RN metric of the mass $m_0$ and the (magnetic)
charge $P^2=V_0^2$, $V_0=V(W_0)$. 
By substituting (\ref{rn}) into (\ref{ep}) one finds that the YM ``potential''
term $V^2$ is dominant, which results in an effective equation of state in the
singularity $\epsilon = p_t = - p_r \sim V_0^2/(2 r^4)$.

The second power-law solution corresponds to the spacelike
(S--type) singularity. In this case the YM field takes a vacuum
value $|W_0|=1$:
\[
 W = W_0 + b r^2 + O(r^5), \quad  m = m_0 - 4 m_0 b^2 r^2 + O(r^3),
\]
where $m_0 \ne 0$ and $b$ are the only free parameters.  This singularity
is dominated by kinetic terms ($W'$). Thus the equation of state is
$\epsilon =p_r \sim 8 b m_0 r^{-3}$, while the tangential pressure tends 
to a constant and thus becomes negligible.

The last power-law solution discovered in~\cite{dgz} corresponds to
the isotropic ``hot'' equation of state $p_r = p_t = \epsilon/3$,
with $\epsilon$ diverging as $r^{-4}$. Here both kinetic and potential
terms give contributions of the same order:
\begin{equation}\label{new} 
 W = W_0 \pm r - W_0 r^2/(2 V_0) + O(r^3), \quad
 \Delta = - V_0^2 \mp 4 W_0 V_0 r + O(r^2),
\end{equation}
(here $W_0 \ne \pm 1$ is the only free parameter). This geometry, conformal
to $R^2 \times S^2$, was encountered in the previous study of black hole
interiors in the framework of the perturbed Einstein--Maxwell theory
(Page, Ori, 1991) and called HMI (homogeneous mass-inflation).

\begin{figure}[t]
\centerline{\epsfxsize=6.5cm \epsfbox{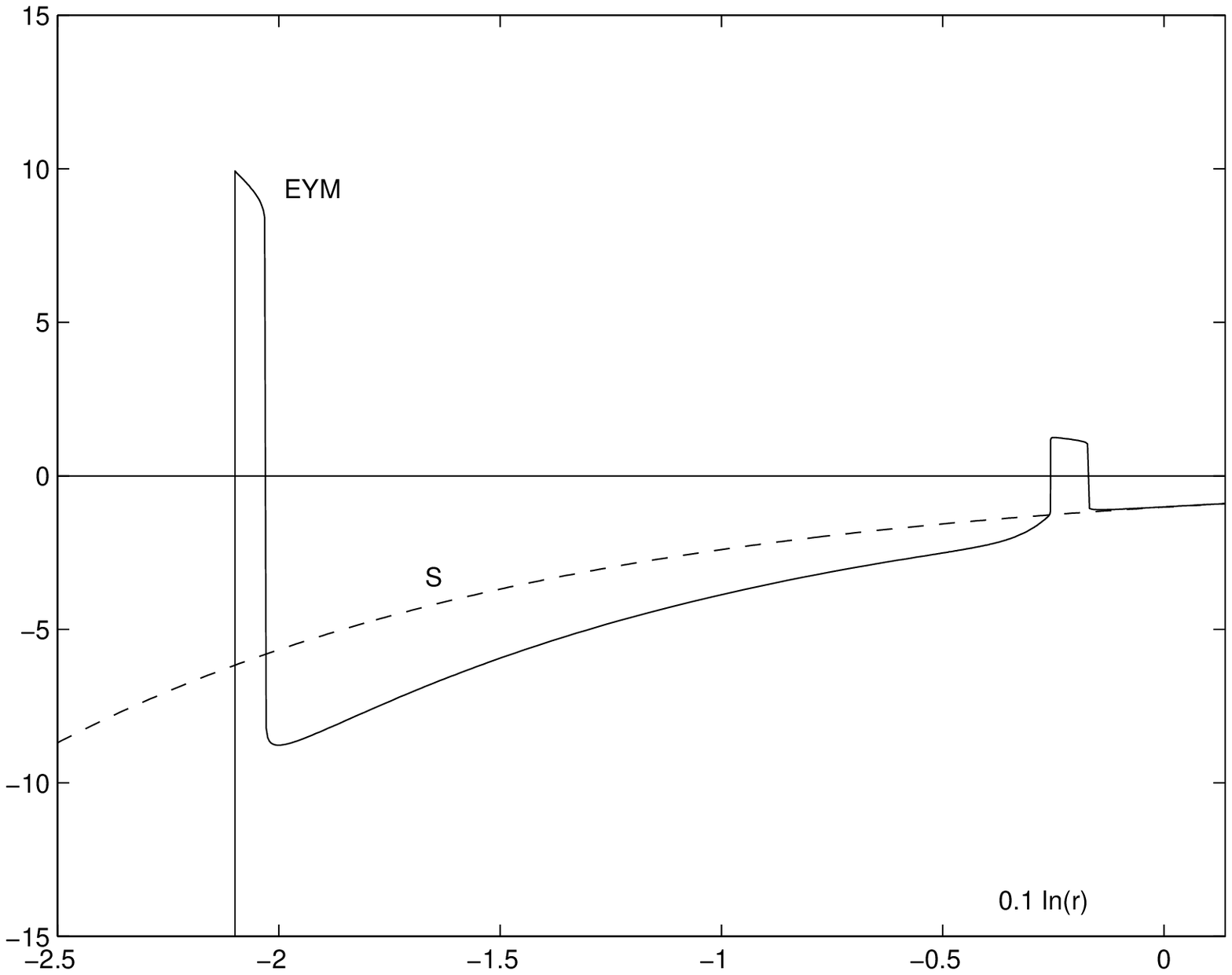}}
\end{figure}
 
Neither of the above singularities may correspond to a {\it generic\/}
black hole. The locally generic RN-type solution (\ref{rn}) is not globally
generic, and is realized only for discrete values of mass.\cite{dgz}
A generic solution does not admit the power-law expansions, but can
be described by a dynamical system
\[
 {\dot q} = p, \quad
 {\dot p} = (3 e^{-q} - 1) p + 2 e^{-2 q} - 1/2,
\]
where $\Delta=-(V_0^2/2) \exp(q)$,
and a dot stands for derivatives with respect to $\tau=2 \ln (r_h/r)$.
The fixed point ($p=0$, $q=\ln 2$) with eigenvalues 
$\lambda=(1 \pm i \sqrt{15})/4$ corresponds to the HMI solution (\ref{new}).
The phase trajectories spiral outward from this point exhibiting
oscillations of $\Delta$ in the negative half-plane with an infinitely
growing amplitude. No Cauchy horizon is met, and the singularity 
is spacelike in conformity with the SCC principle. The Weyl scalar
oscillates, periodically changing sign ($\Psi_2^{1/35}$ for the EYM BH
and S cases ($r_h=4$) is shown in the figure).

When scalar fields are present (we have studied doublet and triplet 
Higgs~\cite{gd} and dilaton~\cite{gdz}), neither $p_r$ nor $p_t$ 
has definite sign, and the stress tensor is not traceless. Meanwhile
the attractor solution near the singularity is described by power law 
and has a very simple form in all cases:
\[
 m = \frac{m_0}{r^{- \lambda^2}}, \qquad
 \sigma = \sigma _1 r^{\lambda^2},
\]
where $\lambda$ is a parameter depending on the black hole mass.
The singularity is spacelike, the Weyl tensor projection is monotonically
diverging.

We conclude that a large class of non-Abelian gravity coupled
theories supports the SCC conjecture by the argument of the generic solutions
behavior.

D.V.G. is grateful to the Organizing Committee for support during
the conference. The work was supported in part 
by the RFBR grants 96-02-18899, 18126.

\end{document}